\documentclass[aps,prd,eqsecnum,showpacs,onecolumn,nofootinbib,preprintnumbers,a4paper]{revtex4}
\input epsf.tex
\epsfclipon

\usepackage{epsfig}
\usepackage{epstopdf}
\usepackage{epsf}
\input{epsf.sty}
\usepackage{graphicx,amsmath,amssymb}
\allowdisplaybreaks

\usepackage{bbm,bm,amsmath,amssymb}
\def\blfootnote{\xdef\@thefnmark{}\@footnotetext}

\long\def\symbolfootnote[#1]#2{\begingroup%
\def\thefootnote{\fnsymbol{footnote}}\footnote[#1]{#2}\endgroup}

\newcommand{\be}{\begin{eqnarray}}
\newcommand{\ee}{\end{eqnarray}}
\newcommand{\ben}{\begin{eqnarray*}}
\newcommand{\een}{\end{eqnarray*}}

\newcommand{\bcent}{\begin{center}}
\newcommand{\ecent}{\end{center}}
\newcommand{\benum}{\begin{enumerate}}
\newcommand{\eenum}{\end{enumerate}}
\newcommand{\bdesc}{\begin{description}}
\newcommand{\edesc}{\end{description}}
\newcommand{\bitem}{\begin{itemize}}
\newcommand{\eitem}{\end{itemize}}
\newcommand{\bquote}{\begin{quote}}
\newcommand{\equote}{\end{quote}}
\newcommand{\bhalfp}{\begin{minipage}{0.45\textwidth}}
\newcommand{\ehalfp}{\end{minipage}}
\newcommand{\bhead}{\begin{center}\bf \Large}
\newcommand{\ehead}{\end{center}\bigskip}

\def\be{\begin{equation}}
\def\ee{\end{equation}}
\def\ba{\begin{eqnarray}}
\def\ea{\end{eqnarray}}

\newcommand{\roughly}[1]{\mathrel{\raise.3ex\hbox{$#1$\kern-0.85em
\lower1ex\hbox{$\sim$}}}}

\def\2pi{\left(2\pi\right)}

\def\beq{\begin{equation}}
\def\eeq{\end{equation}}
\def\bg{\begin{eqnarray}}
\def\nd{\end{eqnarray}}
\def\bea{\begin{eqnarray}}
\def\eea{\end{eqnarray}}

\def\D3{\overline{\mbox{D3}}}

\begin{document}

\title{Cosmological UV/IR Divergences and de-Sitter Spacetime}


\author{Wei Xue, Keshav Dasgupta and Robert Brandenberger}

\affiliation{
Physics Department, McGill University,
3600 University Street, Montr{\'e}al QC, Canada H3A 2T8
\footnote{xuewei,keshav,rhb@hep.physics.mcgill.ca}}

\date{February 2011}


\begin{abstract}

We consider one loop graviton corrections to scalar field Green's functions
in the de Sitter phase of an inflationary space-time, a topic relevant to the 
computation of cosmological observables beyond linear order. By embedding 
de-Sitter space into an ultraviolet complete theory such as M-theory we argue 
that the ultraviolet (UV) cutoff of the effective field theory should be taken to 
be fixed in physical coordinates, whereas the infrared (IR) cutoff is expanding as
space expands. In this context, we demonstrate how to implement three 
different regularization schemes $-$ the brute force cutoff regularization, 
dimensional regularization and Pauli-Villars regularization $-$ obtaining the 
same result for the scalar propagator if we use any of the three 
regularization schemes.

\end{abstract}

\maketitle


\section{Introduction}

The study of quantum fields in an expanding universe has become a cornerstone of modern
cosmology. It is believed that the fluctuations of a scalar quantum field induce
the gravitational fluctuations which in an inflationary universe scenario evolve into the
inhomogeneities in the distribution of galaxies and into the anisotropies of the cosmic
microwave background which we observe today. The computation of the
generation and evolution of cosmological perturbations is typically
performed at linear order (see e.g. \cite{MFB} for an overview
of the theory of cosmological perturbations), i.e. at tree level. Recently, however,
there has been a great deal of interest in studying perturbations at higher order
(see e.g. \cite{higher} for recent reviews). There are both phenomenological and
purely theoretical reasons for this interest. On the phenomenological side, an
important issue has been the study of non-Gaussianities induced by the non-linearities
in the underlying field equations (see e.g. \cite{NG} for some recent reviews and
\cite{early} for two early papers on this subject).
On the purely theoretical side, there are various questions. Foremost, there is
the question of divergences which arise in computing perturbatively to higher
orders. There will clearly be ultraviolet (UV) divergences as there are in any
quantum field theory. Furthermore, in a theory which involves massless modes
like gravity there is the possibility of infrared (IR) divergences. In
inflationary cosmology, these divergences were noticed in early work
\cite{Ford} and lead to a linear growth of the coincident point of the two point
function of a scalar field $\phi$ in the de Sitter phase
\be
\langle\phi^2(x) \rangle \, \sim \, H^3 t \, ,
\ee
where $H$ is the Hubble constant and $t$ is physical time. The role
of these divergences for cosmological perturbations was first
discussed in \cite{Sasaki}. They
could have both interesting and also dangerous effects. For one, they could
invalidate the entire perturbative approach (see e.g. \cite{Losic}). On the other
hand, there have been speculations that IR divergences in the gravity wave
sector \cite{WT} or in the sector of scalar metric fluctuations \cite{Abramo}
could lead to a dynamical relaxation mechanism for the cosmological
constant, a mechanism which would leave behind a remnant cosmological constant
of exactly the right magnitude to explain dark energy \cite{RHBrev}.

The interest has focused on studying quantum fields in a de Sitter background
since such a background is a good approximation for the phase of inflationary
expansion in the very early universe. Free quantum fields in non-trivial
gravitational backgrounds have been studied extensively (see e.g. \cite{BD}
for a textbook treatment), and the study of such fields has evolved into a
mature subject. 

In comparison, the study of interacting fields in cosmological backgrounds is
a field still in its infancy. A good understanding of this topic, however, is vital
if we are to compute correlation functions of cosmological fluctuation
variables to higher than tree order. In particular, we are interested in correlation
function of the variable commonly denoted by either
$\zeta$ or $\cal{R}$ \cite{Bardeen, MFB} which represents the curvature
perturbation in comoving gauge. Since General Relativity is intrinsically non-linear, 
non-linearities will be important for the evolution of $\zeta$ 
even if the matter field is a free field. It is
particularly important to consider the interactions of scalar fields (such as $\zeta$) with
gravitons. For early work on the quantum theory of gravitons
in de Sitter space the reader is referred to \cite{WT}. Early work on
interacting quantum fields in de Sitter-like backgrounds see e.g. 
\cite{Mottola,deVega,Woodard,others}. For a mathematical physics
approach to interacting quantum field theory in cosmological 
backgrounds see e.g. \cite{MP}. 

A few years ago, Weinberg \cite{Weinberg1}
wrote a seminal paper studying quantum contributions to correlation
functions of curvature perturbations in de Sitter space
\footnote{See \cite{fully} for fully dimensionally regulated and renormalized
computations involving gravitons in de Sitter.}. 
Using dimensional regularization, and working in the ``in-in formalism"
\cite{inin} he found the one loop result \footnote{Here, $\zeta$ is the
curvature fluctuation in comoving coordinates - see \cite{MFB}.}
\be
\langle\zeta_k^2\rangle \, \sim \, \frac{1}{k^3} \frac{H^4}{M_{pl}^2} ~{\rm log} \left( \frac{k}{\mu} \right) \, ,
\ee
where $k$ is a co-moving wavenumber, $\mu$ is a physical ultraviolet 
renormalization scale, and $H$ is the Hubble constant during the de Sitter 
phase. As expected, there are both ultraviolet  and infrared 
divergences \footnote{There has been a lot of recent work focusing on
whether the infrared divergences are real or not (see e.g. \cite{Seeryrev} for a review
and \cite{IRviews} for a selection of other references). This
is an issue which we will not touch here. Note that there is a close connection
between the IR divergences and stochastic inflation \cite{stochastic}.} . 
The former are removed via renormalization, but the latter
persist and could have an important effect.  

Similar calculations to those of Weinberg were then performed using
a ``brute-force" cutoff regularization \cite{Sloth, Seery, Easther, Dimas} yielding
consistent results. However, in \cite{SZ} the result of Weinberg \cite{Weinberg1}
was questioned on the basis of the fact that
$k$ is a co-moving scale whereas $\mu$ is physical. The authors of
\cite{SZ} found the result
\be
\langle\zeta_k^2\rangle \, \sim \, \frac{1}{k^3} \frac{H^4}{M_{pl}^2}~ {\rm log} \left( \frac{H}{\mu} \right) \, .
\ee
The difference
between the results of \cite{Weinberg1} and \cite{SZ} can be traced
to different assumptions made about the nature of the UV and IR cutoff
scales (see e.g \cite{Sloth,others2} for other recent work). 
Recently, Weinberg \cite{Weinberg2}
has re-considered the problem and now advocates that one should use 
Pauli-Villars regularization instead of dimensional regularization.
However, one should be able to obtain the same result using different
renormalization schemes \footnote{There have also been some attempts
to go beyond a pure loop expansion \cite{beyond}.}.

The purpose of this note is to address the two issues mentioned in the
previous paragraph. First, by putting the scalar quantum field theory
model in the context of an ultraviolet-complete theory, we will be
able to argue that the correct way to impose the cutoffs is to use a
fixed physical ultraviolet cutoff scale but a co-moving infrared cutoff.
Secondly, we will show in terms of a simple example that all three
regularization schemes commonly used - brute-force cutoff,
dimensional regulations, and Pauli-Villars regularization - all
give the same result.

In the following section we will discuss what we can learn about
renormalization issue of quantum fields in de Sitter space
by embedding de Sitter into an ultraviolet-complete theory such
as string or M-theory. From this discussion, we can draw
conclusions on how an effective
quantum field theory of one low-mass scalar field is embedded in
the ultraviolet-complete theory. In particular, this teaches us how to 
impose the ultraviolet cutoff. In Section III 
we review the quantization of fields in de Sitter space-time.
We then turn to the computation of
the Green's function for a massless scalar field in de Sitter space.
In Sections IV - VI we perform the calculation of the one loop corrections
to the Green's function
using, in turn, brute-force cutoff, dimensional regularization, and
Pauli-Villars regularization. We obtain the same results. In the
final section we discuss our results.

\section{Induced Effective Field Theory in Four Space-Time Dimensions from String Theory}

General Relativity is not renormalizable, and its quantization is an outstanding
challenge. Cosmological perturbations (both scalar metric fluctuations
associated with matter perturbations, and gravitational waves)
can be quantized at the linear level \cite{Sasaki0,Mukh},
but in the absence of an ultraviolet complete theory of quantum gravity which
reduces to General Relativity in the low energy limit questions of consistency
of the quantization scheme remain.

By embedding the de Sitter phase of an inflationary universe into
an ultraviolet complete quantum theory of gravity, it is possible to
study how the conventional theory of fields on de Sitter space arises
as a low-energy effective field theory. This, in turn, will help us
justify the cutoff scales which need to be introduced in order to
eliminate UV and IR divergences.

In spite of ``no-go" theorems \cite{Nunes} derived in the 
context of supergravity, it has been possible in the past years
to construct inflationary solutions of string theory, using input
from string theory which goes beyond simple supergravity
(see \cite{stringinflation} for some reviews)
\footnote{One example of an inflationary model arising from type IIB reduction of
M theory \cite{pisin,dhhk,semil} 
is obtained by considering a D3 brane
moving in the presence of a stack of D7 branes. The D7 branes
wrap the extra dimensions which are taken to have the
form of a particular Calabi-Yau manifold. In the four
dimensional effective field theory, a hybrid inflationary
model of D-term type results: the separation between
the D3 brane and the stack of D7 branes yields the
inflaton field, and the scalar field $\chi$ whose condensation
ends inflation can be identified with a D3-D7 string mode that
becomes tachyonic at a critical value of the D3/D7 interbrane
distance. It has been established 
\cite{gvw,sav,kachruone} that fluxes about
the internal dimensions can stabilize all complex structure
moduli at tree level, and non-perturbative effects can stabilize
the radial moduli \cite{bbdp,kklt}. Alternatively, in a heterotic string
theory, string gases winding the extra dimensions \cite{BV,RHBSGrev}
can stabilize all geometric moduli \cite{Patil,RHBKyoto,Watson,Edna}, and 
gaugino condensation can be used to stabilize the dilaton \cite{DFB}. 
In a separate paper we will revisit the geometry of this D3-D7 system, and show
that from an M-theory starting point the geometry of our
four-dimensional world becomes de Sitter space \cite{us}.}.

String theory (M-theory) is defined in 10 (11) space-time
dimensions. In order to make contact with four-dimensional
physics at low energies, it is crucial to compactify the
extra spatial dimensions and to stabilize their sizes and
shapes (which are the geometrical moduli). String theory
also admits branes and fluxes and they lead to more moduli,
i.e. more degrees of freedom which from the four dimensional
point of view act as fields. There are Kaluza-Klein modes of
the higher dimensional fields which once again yield
four dimensional fields. The upshot of this is that string or
M-theory give rise to a vast number of scalar fields in the
four-dimensional effective field theory.

If all of the moduli are successfully stabilized, then only a
very small number of fields remain light, including the
graviton and the scalar field which plays the role of the
inflaton. All other fields obtain a mass which is characterized
by a combination of the compactification scale and the string
scale. We will denote this scale as $M$. 

Since there are an infinite number of scalar 
fields in four-dimension, all interacting with each other, the UV 
theory is highly non-trivial at energy scales larger than $M$. This
scale is the natural cutoff scale of the low energy effective field
theory. This cutoff scale must be fixed (at least at late cosmological 
times) in physical coordinates to avoid time-dependent coupling 
constants at low energies. Thus, we learn that the ultraviolet cutoff
energy scale is fixed in physical coordinates and hence increasing in
comoving coordinates (as used in most works, in particular in 
\cite{SZ}). 

\begin{figure}[htb]\label{RGfields}
\begin{center}
\includegraphics[height=6cm]{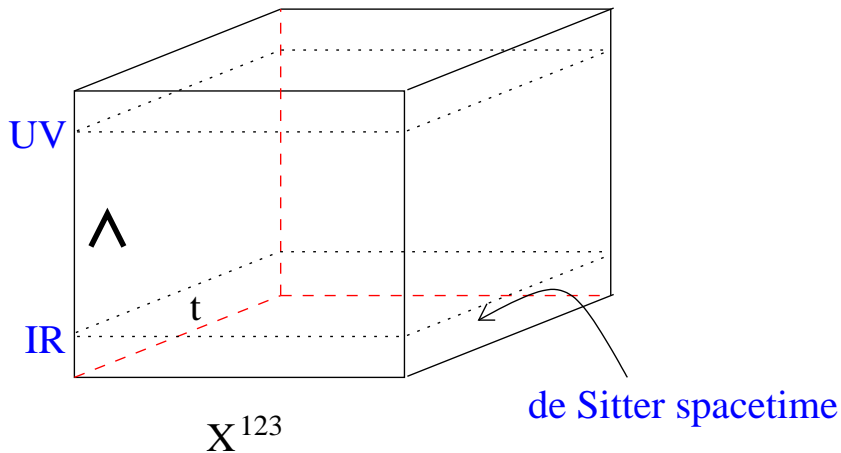}
 \caption{Fields in a de-Sitter space can be arranged succinctly in a 
 five-dimensional space. The horizontal plane represents our four
space-time dimensional expanding universe (the spatial dimensions
in the plane of the paper, the time direction into the paper), and the 
fifth dimension is the energy scale ${\bf \Lambda}\equiv ({\bf \Lambda}_{\rm UV}, {\bf \Lambda}_{\rm IR})$ of the field. 
As the universe expands, the energy of a mode decreases. Thus, as
$t$ increases, the mode moves downwards in the cube. The modes
tracked in the effective field theory lie between the UV and IR cutoffs.
If the z-axis represents physical energy, then the UV cutoff is at a height
which is fixed in time, whereas the IR cutoff scale is decreasing as $t$
increases. New modes continually enter the region of the effective field theory
from the UV sea as $t$ increases.}
\end{center}
\end{figure} 

If we follow a mode with initial frequency smaller than $M$ forwards
in time as space expands, its physical frequency decreases while the
comoving frequency stays the same. If we imagine setting initial
conditions for all modes at the beginning of the inflationary period,
or at some fixed time in de Sitter space (using the cosmological
slicing of de Sitter space), then there is a natural infrared cutoff,
as can be seen by the following argument:
Physics inside the initial Hubble radius $H^{-1}$ cannot determine the initial
conditions on scales larger than $H^{-1}$. Hence, the initial conditions
on initial super-Hubble scales depend on the pre-history of the
cosmological model, the history before the onset on the inflationary
phase. If we are interested in effects due to the de Sitter phase, we must
cut out the modes which are initially super-Hubble. Hence, the
initial Hubble radius is the natural infrared wavelength cutoff.
The wavelengths of modes are stretched as space expands, and hence
the IR cutoff length will expand as well. The IR energy cutoff is fixed in comoving
coordinates but decreasing in physical coordinates.

There is a geometrical picture which represents the setup we
are considering here and which is illustrated in Figure 1. The
horizontal plane represents our four space-time dimensional
universe, the vertical axis is the energy scale. Each mode
corresponds to a wave in the horizontal plane at a particular
height. Modes which lie between the UV and IR cutoff
scales are within the domain of the effective field theory.
As the universe expands, the height of the wave
decreases, and new waves enter the region of the
effective field theory from the UV sea of modes. The IR
cutoff scale decreases as $t$ increases \footnote{Note that
we are not the first to use these IR and UV cutoff prescriptions -
see e.g. \cite{Giddings}.}.

The low energy effective action can be written completely in terms of a
small number of low mass scalar fields
$\sigma_i$ interacting with graviton fluctuations $h_{ij}$. The masses
$m_i$ of all stringy modes and other moduli fields
must be larger than the UV cutoff scale.  Otherwise, the effective action would 
not make any sense beyond scale $m_i$ since these
modes would have to be inserted in the loop diagrams of correlation 
functions of the light modes, e.g. of $\sigma$, which would change the 
behavior of the $\langle\sigma\sigma\rangle$ 
correlator. 

The massive modes contain
the KK reductions of all the ten-dimensional light states in four-dimensions 
over any compactifying manifold as well as the massive stringy modes and 
their KK reductions over the same six-dimensional internal manifold.
All the massive states do couple to gravity, and correspondingly every
low energy state has an infinite number of interaction terms. Thus, with all the 
massive modes and the infinite series expansion of the metric, 
the loop integrals for correlation functions of low mass fields at energies above
the scale $M$ are completely out of control. However as is well 
known, the new stringy states that 
enter the Wilsonian action at the scale where Einstein gravity breaks down change 
the UV picture completely and in fact  control the UV divergences. 
The final UV behavior is finite and show no divergences. 

Thus, it makes sense to study the UV behavior of a single light 
state coupled to gravitational fluctuations up to a specified UV cut-off using the
four-dimensional effective field theory of the low energy modes. Beyond the UV
cutoff we expect the behavior that we discussed above will kick in. 

But for an effective theory to make sense below ${\bf \Lambda}_{\rm UV}$ 
we should be able to regularize the loop integrals unambiguously 
{without resorting to extra massive degrees of freedom}. We must therefore
study the effects of the choice of regulators and of the regularization schemes. 
These questions make sense within the effective four-dimensional action and
will concern the rest of this article.

The simplified Lagrangian of the low energy effective theory, expressed using
the Arnowittt-Deser-Misner (ADM) decomposition of the four-dimensional
space-time metric into spatial metric $g_{ij}$ (the Latin indices run over
the three spatial coordinates), shift vector $N_i$ and lapse function
$N$, is
\bg\label{chaluram}
{\cal L} = \frac{1}{2} \int \sqrt{{\rm det} g_{ij}} \Big[  N R^{(3)} + N^{-1} \left(E_{ij} E^{ij} -E^2 \right)  
+ N^{-1} \left(\dot{\sigma} - N^i \partial_i \sigma  \right) - N g^{ij} \partial_i \sigma  \partial_j 
\sigma\Big]\nonumber \, ,\\   
\nd
where $E_{ij}$ represents the extrinsic curvature tensor, and $R^{3}$ denotes the Ricci
scalar of the spatial metric. We can even insert a potential 
$-2NV$ for the scalar field $\sigma$. The above Lagrangian implies that there are at 
least two interaction vertices:
\bg\label{2vert}
{\cal L}_1~\equiv~ \frac{a}{2} h_{ij} \partial_i \sigma \partial_j \sigma, ~~~~~~
{\cal L}_2 ~\equiv~ -\frac{a}{4}   h_{il} h_{lj} \partial_i \sigma \partial_j \sigma
\nd
which enter into the computation of one-loop corrections of gravitons to 
$\langle\sigma\sigma\rangle$, i.e the two-point correlation function. 

In our five-dimensional picture, \eqref{2vert} will give rise to two distinct non-bifurcating 
surfaces. An alternative, but equivalent, viewpoint will be to take \eqref{chaluram} as the 
Wilsonian action at the scale 
${\bf \Lambda}_{\rm UV}$, and as we go down the scale we will only consider modes with 
momenta up to that scale. Clearly, the
existence of any extra massive states will spoil this simple Wilsonian picture. 

\section{Quantization and Modes in de Sitter Spacetime}

To study the modes in de-Sitter space it is convenient to go to the co-moving 
frame in which the four-dimensional background metric becomes
\bg\label{ghosha} 
ds^2 = a^2(\eta)~\left(-d\eta^2 + \sum_{i = 1}^3 ~dz_i \, .
dz_i\right) 
\nd
For the sake of simplicity, we consider the spatial sections to be flat.
Note that $\eta$ is conformal time.

In the presence of scalar field matter, the metric contains 10 degrees of freedom for
fluctuations, four of them scalar, four vector and two tensor (classifying
the fluctuations according to how they transform under spatial rotations,
the usual procedure in cosmology). Two scalar and two vector modes
are gauge. We fix the gauge by setting four metric fluctuation variables
to zero. Note that in this gauge there are no propagating ghosts \footnote{We thank
Alex Maloney and Guy Moore for asking probing questions which stimulated
us to focus on this issue.} or interactions, unlike
what would happen if one were to use a covariant gauge-fixing procedure
(see e.g. \cite{gauge} for studies using covariant gauges). The Hamiltonian
and momentum constraints eliminate four further variables, leaving us
with one \footnote{For $n$ scalar fields there would be $n$ scalar modes.}
scalar mode (a combination of the scalar matter field and the
scalar metric fluctuation) and the two tensor modes which are the two
graviton polarization fields. The scalar
field and the graviton obey the following differential equations (with a prime
denoting the derivative with respect to conformal time):
\bg\label{mayah} 
&&{\sigma}'' + 2 a {H}  {\sigma}' - \triangledown^2 \sigma ~= ~ 0 \nonumber\\
&&{h}_{ij}'' + 2 a {H} {h}_{ij}' - \triangledown^2 h_{ij}~= ~0
\nd
We can use the above equations to describe the modes of the scalar field and the metric. 
The mode expansion for the scalar field is the standard one. For the graviton we will use 
a basis of polarisation tensors $e_{ij}(k, s)$ ($s = \pm$) to describe the mode expansion as:
\bg\label{gravmodes} 
h_{ij} ({\bf z}, \eta)= \int d^3 k \sum_{s=\pm} \Big[  a({\bf k}) e_{ij}(k, s) h_{s, k}(\eta) 
e^{i{\bf k} \cdot {\bf z}} +a^\dagger({\bf k}) {e_{ij}}^*(k, s) h_{s, k}^*(t) e^{-i{\bf k} \cdot {\bf z}}\Big] \, ,
\nd
where $h_{\pm, k}$ denote the  mode functions. The polarisation tensors satisfy the identity:
\bg\label{polT} 
\sum_s e_{ij}(\hat p,s) e_{kl}^* (\hat p, s) =&& ~~\delta_{ik}\delta_{jl}~+~\delta_{il}\delta_{kl}~-~\delta_{ij}
\delta_{kl}
~+~ \delta_{kl}\hat p_i\hat 
p_j~-~\delta_{ik}\hat p_j\hat p_l\nonumber\\
&&~~-~\delta_{il}\hat p_j\hat p_k ~-~\delta_{jk}\hat p_i\hat p_l
~-~\delta_{jl}\hat p_i\hat p_k+\hat p_i\hat p_j\hat p_k\hat p_l \, ,
\nd
and the subscript $s$ of the graviton is the helicity. We have used $\hat p$ to denote the 
unit momenta and $k$ to denote the absolute value. 

The mode functions in de Sitter space are easy to obtain. We assume that the modes
start out on sub-Hubble scales in their vacuum state, the usual assumption made in
inflationary cosmology. Let us denote the scalar field modes that appear in the standard mode 
expansion by $u_k$. The final result for these modes is:
\bg\label{modeli}
&&u_k ~ = ~ \frac{H}{(2\pi)^{3/2}\sqrt{2k^3}} \left( 1+i k\eta \right) e^{-ik\eta}\nonumber\\
&& h_{\pm, k}~ = ~ \frac{H}{(2\pi)^{3/2}\sqrt{k^3}} \left( 1+i k \eta\right) e^{-ik\eta} \, ,
\nd
where we note that both the modes $h_{+, k}$ and $h_{-, k}$ are given by the same expression. 

Now that we have identified the physical modes of the system, we can move on
to study the behavior of these modes in a de-Sitter phase of an inflationary
universe. Specifically, we need to understand what it means to impose IR and UV 
cut-offs in an expanding universe. The mode decomposition of the fields
is defined in comoving coordinates. As we have argued in the previous section,
the UV cutoff of the effective field theory should be at a fixed physical wavelength,
and hence at a comoving wavelength which is decreasing as $a(t)^{-1}$. The
comoving high frequency cutoff is hence increasing. In physical coordinates,
it is the UV cutoff which remains the same whereas the physical IR wavelength
cutoff is increasing. Since the Hubble radius $H^{-1}(t)$, the length which separates
the high frequency region where the modes are oscillating from the low frequency
region where the modes are frozen out, is at a fixed value in physical coordinates,
we conclude that the volume of phase space of short wavelength modes is constant, whereas
that of the long wavelength modes is increasing. As short wavelength modes exit
the Hubble radius during the inflationary expansion of space, new UV modes enter
the region of validity of the effective field theory to replenish the phase
space.

\begin{figure}[htb]\label{modesindesit}
\begin{center}
\includegraphics[height=6cm]{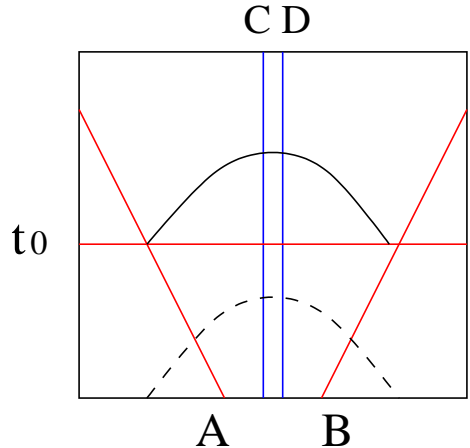}
\caption{A representation of how modes evolve in de Sitter space. The vertical
axis denotes time, the horizontal axis physical space. The grey dashed half wave and
the solid grey half wave at time $t_0$ denotes waves of Hubble length. The
red slanted solid lines delineate the wavelength of a wave which exits the
Hubble radius at time $t_0$ (whose initial value is given by the length
between A and B). The solid blue vertical lines indicate the ultraviolet
cutoff wavelength (the distance between C and D) which is constant in physical 
coordinates. }
\end{center}
\end{figure}

{F}igure 2 gives a sketch of the evolution of various scales in de Sitter space-time.
The horizontal axis represents physical distance, the vertical distance time.
The wavelength of a fixed mode increases as time proceeds. If $t_0$ is
the beginning of inflation, then the mode whose wavelength is indicated by
the distance between the two red solid slanted lines indicates the IR cutoff
scale whose wavelength is growing with time. In constrast, the UV cutoff
wavelength is fixed in physical coordinates - by the distance between C and D
in the sketch. The Hubble radius has constant physical distance. 

In the following, we will consider the one-loop graviton corrections to the two point 
function of $\sigma$. We perform the computation using three different 
regularization schemes:

\vskip.1in

\noindent $\bullet$ Brute cut-off regularization,

\vskip.1in

\noindent $\bullet$ Dimensional regularization,

\vskip.1in

\noindent $\bullet$ Pauli-Villars regularization.

\vskip.1in

\noindent and show that all the three regularization schemes yield identical results\footnote{One might worry about 
preserving the Ward identity using the above regulators. The Pauli-Villars regulator (as like the 
dimensional regulator) {\it does} preserve the Ward identity. However,
to preserve the Ward identity using the brute cut-off 
regulator one might need to add extra terms to the lagrangian. The final answer is that any violation of the Ward 
identity is cancelled by these extra terms {\it without} affecting the two-point correlation functions. 
Therefore in this work
we will ignore these subtleties.}.   

In an upcoming work \cite{toappear2} other interactions of $\sigma$ will be
studied, and an extension of our method to higher $n$-correlation functions will be 
worked out. However, we want to point out that the steps presented in this paper
to regularize the one-loop two-point function extend in a straightforward way to 
higher $n$-point functions, including non-trivial fermionic and gauge interactions.  

\section{Brute Cut-off Regularization}

\subsection{Preliminaries}

The ``Brute Cut-off'' scheme consists of eliminating the contributions of modes
with wavenumbers above a certain ultraviolet cutoff scale and below a certain
infrared cutoff scale from the loop integrals. The specification of these
cutoffs is the first place where all the subtleties that we mentioned above 
regarding UV and IR cut-offs will show up. 

In scattering experiments one is interested in computing transition rates between
prescribed in and out states. The questions in cosmology are very different.
Here, one is interested in the evolution for a finite duration of time of a
certain initial state. Thus, we are not interested in calculating S-matrix
elements, but rather expectation values of operators at some later time $t$
evaluated in a state set $\Omega$ up at some initial time $t_i$. The formalism
appropriate for performing this computation is the Schwinger-Keldysh \cite{inin}
or ``in-in'' formalism.

We are interested in computing the expectation value
\be \label{corfct}
\left<\Omega| \mathcal{O} (t)   |\Omega  \right>
\ee
of some operator $\mathcal{O}$ evaluated at time $t \gg t_i$ in some state $\Omega$
prescribed at some early time $t_i$ which we consider to tend to $- \infty$. We
take the state $\Omega$ to be the initial vacuum state. 
Working in the interaction representation, the formula for this correlation
function (\ref{corfct}) becomes
\bg\label{inin} 
\left<\Omega| \mathcal{O} (t)   |\Omega  \right> = 
\langle 0 \vert\bar{T}e^{i \int_{-\infty}^{t}H_I(t') dt'}  
{\cal O}_I (t)   {T}e^{-i \int_{-\infty}^{t}H_I(t') dt'} \vert 0 \rangle
\nd
where $\vert\Omega \rangle$ is the vacuum in the interacting theory and 
$\vert 0 \rangle $ is the free field vacuum. The right hand side of the 
equation is calculated in the interaction picture. 
$T$ and $\bar{T}$ denote time-ordered and anti-time-ordered products.

Considering a free scalar field theory for matter \footnote{The be complete,
we should consider the self-coupling of $\sigma$ which is unavoidable if
the coupling of matter to scalar gravitational fluctuations is
taken into account.}, there are two diagrams which contribute to the
above expectation value at the one loop level. They are shown in 
Figure 3. 
\begin{figure}[htb]\label{feynman1}
\begin{center}
\includegraphics[height= 5cm]{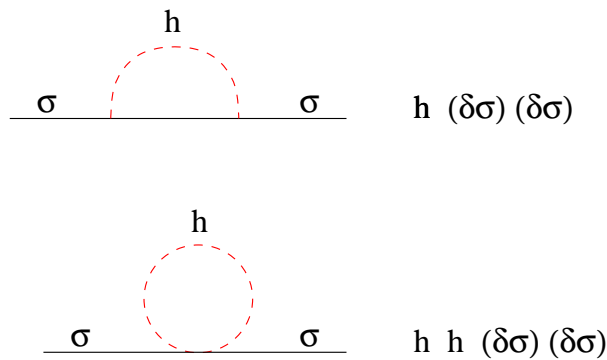}
\caption{The two one-loop diagrams which we study here. They are related to 
   the two amplitudes given in the text. The first diagram gives amplitude I 
   and the second one amplitude II.}
\end{center}
\end{figure}
We will evalute these diagrams in the following subsection.

\subsection{In-in calculation of loop corrections to the two point function} 

The contribution $G_p^{I}(\eta)$ of the first diagram of Figure 3 to the  
one loop corrections of the scalar correlation is due to the three field 
interaction \eqref{2vert} is given  
by the following expression:
\bg\label{Gp1_c}
G_p^{I}(\eta) &=& -4 (2\pi)^6 {\bf Re} \int_{-\infty}^{\eta}d\eta_1 a^2(\eta_1)  
\int_{-\infty}^{\eta}d\eta_2 a^2(\eta_2) 
\int d^3 q\  p^4 {\rm sin}^4 \theta \nonumber\\
 &&\times \Big[\theta (\eta_1-\eta_2) u_p^2(\eta) u_p^*(\eta_1) u_p^*(\eta_2)  u_{p'}(\eta_1) u_{p'}^*(\eta_2) 
h_q(\eta_1) h_q^*(\eta_2)\nonumber\\
 &&-\frac{1}{2}  |u_p(\eta)|^2  u_p(\eta_1) u_p^*(\eta_2)  u_{p'}(\eta_1) u_{p'}^*(\eta_2) h_q(\eta_1) 
h_q^*(\eta_2)\Big] \, ,
\nd
where $p$ is the momentum of the external line, $q$ is the momentum of the graviton
and $p' \equiv p - q$ is the momentum of the internal
scalar field. The factor $p^4 {\rm sin}^4 \theta$ comes from summing over the 
graviton polarization states:
\bg\label{mtan}
\sum_s e_{ij}({\hat p},s) e_{kl}^* ({\hat p}, s) p_i p_j p_k p_l ~=~ p^4 {\rm sin}^4 \theta \, .
\nd

Note that in Eq. \eqref{Gp1_c}, the second part in the bracket is exponentially small 
in the UV because of the exponential oscillation of the modes. Therefore, when we 
calculate the UV divergence of Diagram I, we can 
just integrate the first part without effecting the result. In the UV limit, the 
correlation can then be simplified as:
\bg\label{Gp1_UV}
G_p^{I, UV}(\eta) &=& -4 {\bf Re}   \int_{-\infty}^{\eta}d\eta_1  \int_{-\infty}^{\eta_1}d\eta_2   u_p^2(\eta) u_p^*(\eta_1) u_p^*(\eta_2)   \int \frac{d^3 q }{2q^2}  p^4 {\rm sin}^4 \theta e^{ -2i q \left( \eta_1 -\eta_2 \right)  }  \nonumber\\
 & = & - ~{}^{\rm lim}_{\eta_1 \to \eta_2} \int_{-\infty}^{\eta}d \eta_1 {\bf Im} \left[  u_p^2(\eta) {u_p^*}^2(\eta_1)  \right] \int \frac{d^3 q}{q^3} p^4 {\rm sin}^4 \theta \, .
\nd
On the other hand, if we want to keep both the IR and UV effects, we can take \eqref{Gp1_c},
do the time integral first and then integrate over the loop momentum. 
The final result (writing only the divergent part) can be expressed 
in the following way:
\bg\label{Iloop}
G_p^{I}(\eta=0) ~= ~ \frac{H^2}{(2\pi)^3 2 p^3} \frac{H^2}{(2\pi)^2} \left[  2\log (p/\Lambda_{IR})  +\frac{2}{3} \log (\Lambda_{UV}/p) \right] 
\nd
where $H$ is the Hubble constant and $p$ is the incoming momentum of the scalar field. 

Once we have the result for Diagram I of Figure 3, we can go to evaluate 
Diagram II. This follows by arguments more or less similar to those made
when considering in \eqref{Gp1_c}. The one loop correction to the scalar 
correlation function from the four field interaction in \eqref{2vert} then
takes the form:
\bg\label{digra2}
G_p^{II}(\eta) &=&  (2\pi)^3 \int_{-\infty}^{\eta}d \eta_1 a^2(\eta_1) {\bf Im} \left[u_p^2(\eta) {u_p^*}^2(\eta_1) 
\right] 
\int d^3 q \ 2 p^2 {\rm sin}^2 \theta |h_q(\eta_1)|^2
\nd
where now the quantity $2p^2 {\rm sin}^2 \theta$ comes from the summation of 
the following graviton polarization:
\bg
\sum_s e_{il}({\hat p},s) e_{jl}^* ({\hat p}, s) p_i p_j =2 p^2 {\rm sin}^2 \theta \, .
\nd
To evaluate this one should once again first do the time integral and then 
integrate the loop momentum. The final result (once again writing only the 
divergent terms) is given by:
\bg\label{digII}  
 G_p^{II}(\eta=0) ~= ~ \frac{H^2}{(2\pi)^3 2 p^3} \frac{H^2}{(2\pi)^2} \left[ 2 \log (\Lambda_{IR} / \Lambda_{UV}) 
-\frac{5}{6} \Lambda_{UV}^2 \right] \, ,
\nd
where all the terms appearing here have been defined above. 
So far, the two results \eqref{Iloop} and 
\eqref{digII} agree with those given in \cite{Giddings} (see also \cite{SZ}). 

Finally, we can rewrite \eqref{Iloop} and \eqref{digII} using the renormalisation
scales $\mu_{IR}$ and $\mu_{UV}$. In this language the two amplitudes are re-expressed as: 
\bg\label{twodim} 
G_p^{I}(\eta=0) &= & \frac{H^2}{(2\pi)^3 2 p^3} \frac{H^2}{(2\pi)^2} \left[  2\log (p/\mu_{IR})  +\frac{2}{3} \log (\mu_{UV}/p) \right]\nonumber\\ 
 G_p^{II}(\eta=0)& =& \frac{H^2}{(2\pi)^3 2 p^3} \frac{H^2}{(2\pi)^2} \left[ 2 \log (\mu_{IR} / \mu_{UV}) -\frac{5}{6} \mu_{UV}^2 \right] \, .
\nd
However, this is {\it not} the final answer because there is a subtlety 
associated with applying cut-offs in the co-moving and in the physical 
coordinate systems. To understand this and analyse the results correctly, 
we turn to the implementation of the cutoffs. 

\subsection{Physical UV and comoving IR cutoffs}

In the previous subsection we have not carefully discussed the 
UV and IR cut-offs for our case. The subtlety about this has been 
discussed earlier in Section 2. 
It is now time to apply what we learned in that section 
to our two results \eqref{Iloop} and \eqref{digII} or 
equivalently to \eqref{twodim}. 

Note that the calculations in the above subsection are based on the assumption that 
the UV and IR cut-offs are both co-moving. During inflation, we know that the 
UV modes are generated and IR modes are stretched outside of the Hubble radius. 
Thus, the phase space is increasing in de Sitter space. Therefore (using
for the moment a superscript ``c'' to designate a quantity in comoving coordinates), 
in the co-moving 
frame the IR cut-off $\Lambda^c_{IR}$ will remain unchanged, whereas the UV cut-off 
$\Lambda^c_{UV}$ will change because new high energy modes will enter the 
system. Therefore we can rewrite $\Lambda^c_{UV}$ as 
\be
\Lambda^c_{UV} \, = \, \lambda_{UV} a(\eta)
\ee
such that $\lambda_{UV}$ is the physical UV cut-off (i.e which remains unchanged 
in the physical ($x^i, t$) frame). Thus, we can re-express Eq. \eqref{digra2} as:
\bg\label{fmchu}
G_p^{II}(\eta)~ =~ (2\pi)^3 \int_{-\infty}^{\eta}d \eta_1 a^2(\eta_1) {\bf Im} \left[u_p^2(\eta) {u_p^*}^2(\eta_1) \right] 
\int^{\lambda_{UV} a(\eta_1) }_{\Lambda_{IR}} d^3 q \ 2 p^2 {\rm sin}^2 \theta |h_q(\eta_1)|^2\nonumber\\
\nd
where both $\lambda_{UV}$ and $\Lambda_{IR}$ are constant. Thus, completing the 
above integral and replacing the cut-offs by the renormalisation scales 
${\widetilde \mu}_{UV}$ and $\mu_{IR}$, we get:
\bg\label{nhart}  
 G_p^{II}(\eta=0) \, = \, \frac{H^2}{(2\pi)^3 2 p^3} \frac{H^2}{(2\pi)^2} \left[ 2 \log (\mu_{IR} / p) 
+ 2\log (H/{\widetilde \mu}_{UV}) - \frac{{\widetilde \mu}_{UV}^2}{H^2} \right] \, .
\nd
For the other one-loop diagram in Figure 3 i.e
$G_p^{I}$, we can use \eqref{Gp1_c} and \eqref{Gp1_UV} to easily obtain the following result:
\bg\label{atexs} 
G_p^{I}(\eta=0) =\frac{H^2}{(2\pi)^3 2 p^3} \frac{H^2}{(2\pi)^2} \left[2\log (p/\mu_{IR})  
+\frac{2}{3} \log ({\widetilde\mu}_{UV}/H) \right] 
\nd
where one may note the appearence of $H$ inside the logarithm as above. 
Note also that ${\widetilde\mu}_{UV}$ is the physical renormalization scale, 
whereas $\mu_{IR}$ is the comoving renormalisation scale.

\section{Dimensional regularization}

Our next step is to analyse the dimensional regularization of the two one-loop 
diagrams of Figure 3. Dimensional regularization is rather subtle here because 
we need to first find the modes in de-Sitter space in 
$d = 4 + \delta$ dimensions where $\delta \to 0$. This is related to one of 
the issues associated with dimensional regularization: absence of the usual 
analytic form of the action integrand as a function of the wave number. Of 
course a way out of this problem is to analyse the system in a slow roll 
inflationary scenario or an equivalent de-Sitter spacetime. 

In the following we will start by finding the modes in $d$ spacetime dimensions 
for the de-Sitter background. 

\subsection{Free field quantization}

In $d$ dimensional spacetime, the Lagrangian of the massless scalar field in 
the conformal FRW background metric is written as
\be 
\mathcal{L}= \sqrt{-{\rm det}~ g} \left[ -\frac{1}{2} a^{-2}\eta^{\mu\nu} \partial_\mu \sigma \partial_\nu \sigma  \right] = -\frac{1}{2} a^{d-2}\eta^{\mu\nu} \partial_\mu \sigma \partial_\nu \sigma  \, .
\label{Ld}
\ee
Canonical quantization requires that the field should be redefined. 
Considering the power of the scale factor $a$ in Eq. \eqref{Ld}, we are
led to defining canonical variables $v$ and ${\widetilde{h}}$ as
\bea
v \, &\equiv& \, \sigma a^{-1-\delta/2} \nonumber \\
\widetilde{h} \, &\equiv& \, \frac{h}{\sqrt{2}}a^{-1-\delta /2} \, , 
\eea
where $\delta = d-4$.
In the limit $\eta \rightarrow -\infty$, we get:
\bg\label{avade} 
v_k ~&= &~ \frac{1}{(2\pi)^{3/2}\sqrt{2k}} e^{-ik\eta}\nonumber\\
\widetilde{h}_{\pm, k} ~&= & ~ \frac{1}{(2\pi)^{3/2}\sqrt{2k}} e^{-ik\eta}
\nd

Once we know the behavior for the redefined components at $\eta \to -\infty$, 
one may use them to determine the expressions for the modes $u_k$ and $h_{\pm, k}$.  
With the usual vacuum initial conditions \eqref{avade}, the $d$-dimensional 
equation of motion of the free fields can be solved 
to give us the following mode expansions: 
\bg\label{keldev} 
u_k~&=&~\frac{e^{i\pi (1+\frac{\delta}{4})}H^{1+\frac{\delta}{2}}\left(   -k \eta    \right)^{\frac{3+\delta}{2}}}
{ 4\pi \sqrt{2} k^{\frac{3+\delta}{2}}}   {\cal H}_{\frac{3+\delta}{2}}\left(-k \eta  \right)\nonumber\\
h_{\pm, k} ~ &= & ~\frac{e^{i\pi (1+\frac{\delta}{4})}H^{1+\frac{\delta}{2}}\left(-k \eta\right)^{\frac{3+\delta}{2}}}
{ 4\pi k^{\frac{3+\delta}{2}}} {\cal H}_{\frac{3+\delta}{2}}\left( -k \eta  \right) \, ,
\nd
where ${\cal H}_{\frac{3+\delta}{2}}$ is a Hankel function. 
The $d$-dimensional modes can now be Taylor expanded around 
$\delta = 0$ to give us the following mode components: 
\bg\label{miklee} 
u_k ~=~ \frac{H}{(2\pi)^{3/2}\sqrt{2k^3}} \left( 1+i k \eta \right) e^{-ik\eta} \left(1+ \frac{\delta}{2} \log (-H\eta)   
+ \cdots  \right)
\nd
Note that there is an extra term $\delta \log (-H\eta)$ in the mode expansion. 
It is important to consider this term in dimensional regularization. As was 
pointed out in \cite{SZ}, this correction term will change the 
one-loop final result from $\log(k/\mu)$ to $\log(H/\mu)$. In the following 
analysis we will carefully consider the implication of this term in
the regularization process.

\subsection{Zeroth order regularization}

Let us start by ignoring the $\delta \log (-H\eta)$ term in the mode expansion. 
We will call this the zeroth order calculation and in the next section we will 
introduce the correction term. For dimensional regularization, it is 
useful to introduce the RG scale $\mu_D$ in the interaction lagrangians that 
modifies our earlier interactions \eqref{2vert} in the following way:
\bg\label{modify}
&&{\cal L}_1 = \frac{a\mu_D^{-\frac{\delta}{2}}}{2} h_{ij} \partial_i \sigma \partial_j \sigma\nonumber\\
&&{\cal L}_2 = \frac{a\mu_D^{-\delta}}{2} h_{il} h_{lj} \partial_i \sigma \partial_j \sigma \, .
\nd
Using these, the value of the first diagram in Figure 3
can now be worked out in the following way:
\bg\label{Gp1d}
G_p^{I}(\eta) &=& -4 (2\pi)^{6+2\delta} {\bf Re} \int_{-\infty}^{\eta}d \eta_1 a^{2+\delta}(\eta_1)  
\int_{-\infty}^{\eta}d \eta_2 a^{2+\delta}(\eta_2) \int d^{3+\delta} q\  p^4 {\rm sin}^4 \theta \nonumber\\
 &&\times \left[  \theta (\eta_1-\eta_2) u_p^2(\eta) u_p^*(\eta_1) u_p^*(\eta_2)  u_{p'}(\eta_1) u_{p'}^*(\eta_2) h_q(\eta_1) h_q^*(\eta_2)       \right. \nonumber\\
 && \left. -\frac{1}{2}  |u_p^2(\eta)|^2  u_p(\eta_1) u_p^*(\eta_2)  u_{p'}(\eta_1) u_{p'}^*(\eta_2) h_q(\eta_1) h_q^*(\eta_2)  \right] \, .
 \nd

The analysis of the above integral will be a little easier compared to the integrals
arising in the brute cut-off scheme because in the current scheme the cut-off 
$\delta$ does not dependend on time. This means that we can do the time integral 
first, yielding the following expression: 
\bg\label{timlight}
G_p^{I}(\eta) = \frac{H^4}{(2\pi)^6 p^4} \int d^{3+\delta} q \int d^{3+\delta} p' \delta^3\left(   \textbf{p}+\textbf{p}'+\textbf{q} \right) f\left( p,p',q \right) \, ,
\nd
where we have kept all the momentum dependences from \eqref{Gp1d} in the 
function $f(p, p', q)$. An immediate advantage 
in writing \eqref{Gp1d} in the form \eqref{timlight} is that we can use 
dimensional analysis to predict: 
\bg\label{dimanal} 
\int d^{3+\delta} q \int d^{3+\delta} p' \delta^3\left(   \textbf{p}+\textbf{p}'+\textbf{q} \right) f\left( p,p',q \right)  
~ \propto ~  p^{1+\delta} F \, ,
\nd
where the constant of proportionality will be determined below, and 
$F$ represents the following functional form in either the UV or the IR:
\bg\label{fdef}
F~=~ \frac{F_0}{\delta} +F_1
\nd
where $F_0$ and $F_1$ are both independent of $\delta$ as expected. 
In the above equation \eqref{fdef} one might wonder 
why $F$ just has a single pole. This is because in flat spacetime one 
loop dimensional regularization has only a single pole. The curved 
spacetime cannot yield more poles. 

In the limit of $\delta \rightarrow 0$, the $\delta$ appearing in the 
denominator cancels out in the standard way to 
give us the following result:
\be 
\int d^{3+\delta} q \int d^{3+\delta} p' \delta^3\left({\bf p}+{\bf p}'+{\bf q} \right) f\left( p,p',q \right)  ~=~ 
p (F_0 {\rm log}~ p + {\rm constant}) \, ,
\ee
where the constant term in independent of $p$ but allows a $\delta^{-1}$ 
singular term. This divergence is not a problem as it will be 
renormalised by the UV completion of our theory. 

Let us now choose a momentum scale $p$ such that we can divide the $d^3q$ momentum 
integral into two parts \footnote{This is because $p$ is the only allowed scale 
for us here. Therefore this gives a natural demarkation between UV and IR physics.} 
$0 \le q \le p$ and $p \le q \le \infty$. We can equivalently divide $F_0$ into two 
parts: $F_0^{(IR)}$ and $F_0^{(UV)}$ whose values can be read off from \eqref{Gp1d}. 
This gives us \footnote{It is easy to see that the UV and IR divergence are determined 
by the sign of $\delta$. First consider the momentum integral, which contains UV and 
IR divergences simultaneously:
$$\int_{\Lambda_{IR}}^{\Lambda_{UV}} \frac{d p}{p} = 
{\rm log}~ \left(\frac{\Lambda_{UV}}{\Lambda_{IR}}\right) \, .$$
This integral, which is defined in four dimensions, can be rewritten in $d = 4 +\delta$ 
dimensions by making the following change:
$$\int \frac{d p}{p^{1-\delta}} = \frac{p^\delta}{\delta } \, .$$
The $\delta$ dependence of the above integral tells us that 
when we use dimensional regularization to deal with UV and IR 
divergences, different choices of $\delta$ should be made. For example, 
when $p \rightarrow \infty$, and $\delta < 0$ can keep the UV divergence 
under control. Whereas when $p \to 0$ then $\delta > 0$ in order to keep the 
IR divergence under control.}:
\bg\label{divis}
&&\frac{2\pi F_0^{(UV)}}{3} \int^\infty_p \frac{d q}{q}  \left(\frac{q}{{\mu}_{D, UV}} \right)^\delta = 
p\left[\frac{2\pi F_0^{(UV)}}{3} {\rm log}~ ({\mu}_{D, UV} /p)  
+ {\rm constant} \right]\\
&&2\pi p F_0^{(IR)} \int_0^p \frac{dq}{q}\left(\frac{q^2}{\mu_{D, IR}(\eta_1)\mu_{D, IR}(\eta_2 )} \right)^{\delta/2}
= \Big[a(\eta_1) a(\eta_2)\Big]^{\delta/2}\nonumber\\ 
&&~~~~~~~~~~~~~~~~~~~~~~~~~~~~~~~~~~~~~~~~~~~~~~~~~~~~~~\times p \Big[2\pi F_0^{(IR)}{\rm log}~(a \mu_{D, IR}/p)
+ {\rm constant}\Big] \, ,\nonumber
\nd
where as before, the constant parts in the above equations are 
divergent but independent of $p$. Note also that, since in dimensional 
regularization we always choose physical renormalization group (RG) 
scales, they are related 
to the RG scales used in the brute cut-off scheme of the previous section 
in the following way:
\bg\label{scalemat}
\mu_{D, UV} ~ = ~ {\widetilde\mu}_{UV}, ~~~~~~~ \mu_{D, IR} ~ = ~ {\mu_{IR} \over a} \, .
\nd
Combining \eqref{divis} with \eqref{Gp1d}, we get our final result for 
the zeroth order analysis:
\bg\label{glynn} 
G_p^{I,0}(\eta=0) ~= ~ \frac{H^2}{(2\pi)^3 2 p^3} \frac{H^2}{(2\pi)^2} \left[2\log (p/\mu_{IR}) + 
\frac{2}{3} \log ({\widetilde\mu}_{UV}/p) \right] \, ,
\nd
where the subscript in $G_p^{I,0}$ denotes the one-loop term without 
the $\delta \log (-H\eta)$ correction. Note also that
we have chosen the RG scales $\mu_{IR}$ and $\mu_{UV}$ as before
\footnote{Note that the additional 
$1+ {\delta\over 2} {\rm log}\left(-H\eta\right)$ correction term coming 
from $u^2_p(\eta)$ in \eqref{Gp1d} only 
adds a correction term to \eqref{glynn} that can be absorbed in the 
defination of the renormalisation scale.}. 

For the second diagram in Figure 3 we can basically follow the same 
set of ideas to get the UV and IR divergence.
The amplitude for the second diagram is given by:
\bg\label{ftucci}
G_p^{II}(\eta) ~=~  (2\pi)^{3+\delta} \int_{-\infty}^{\eta}d \eta_1 a^{2+\delta}(\eta_1) {\bf Im} \left[u_p^2(\eta) 
{u_p^*}^2(\eta_1) \right] \int d^{3+\delta} q ~2 p^2 {\rm sin}^2 \theta |h_q(\eta_1)|^2
\, . \nonumber\\
\nd
If we ignore the $\delta \log (-H\eta)$ correction to the modes, we can 
rewrite the amplitude as:
\bg\label{whitstev} 
 G_p^{II,0}(\eta=0) &\simeq & \frac{H^4}{(2\pi)^6 2p^3} \int d^{3+\delta} q  
\frac{\left(6 p^2+5 q^2\right) 2 p^2 {\rm sin}^2 \theta}{8 p^4 q^3}\nonumber\\
 &= & \frac{H^2}{(2\pi)^3 2 p^3} \frac{H^2}{(2\pi)^2} \left[ 2 {\rm log} (\mu_{IR} / p)
+2{\rm log} (p/ {\widetilde\mu}_{UV})  \right] \, .
\nd

\subsection{First order regularization}

Let us now add back the $\delta {\rm log} (-H\eta)$ correction term. 
In the following we will compute the effect of this
addition to our loop regularization. Before we go about doing this, note that
every mode has this kind of correction. In the loop integrals for the 
two diagrams, i.e \eqref{Gp1d} and \eqref{ftucci}, the 
$\delta {\rm log} (-H\eta)$ in the loop can be cancelled by the scale factor
in the equation. This is because 
the correction disappears if the modes are multiplied with $a^{\delta/2}$. 
When the loop diagram is considered, 
there is a factor proportional to  $a^{2+\delta}$ coming from the time integral. 
However the correction from $u^2_p(\eta)$ can be absorbed in the redefination 
of the renormalisation scale, as we mentioned earlier. 

Taking all the above into account, the correction to $G_p^{II}$ is as follows:
\bg\label{corr1} 
 G_p^{II,1}(\eta=0) \simeq \frac{H^4\delta}{(2\pi)^6 2p^3} \int d^{3+\delta} q{\rm log} (H/p)
\frac{\left(6 p^2+5 q^2\right) 2 p^2 {\rm sin}^2\theta}{8 p^4 q^3}+ \delta\times {\rm constant} \, .\nonumber\\
\nd
The above integral can be computed in exactly the way that we did in the previous 
subsection, i.e by dividing into 
two parts dealing with IR and UV divergences respectively. In fact there are no 
IR divergences because
the prefactor $\Big[a(\eta_1) a(\eta_2)\Big]^{\delta\over 2}$ of \eqref{divis}
in the IR calculations will cancel the first order corrections from $u_p(\eta_1) u_p(\eta_2)$ in 
\eqref{Gp1d} and  \eqref{ftucci}. 
Therefore combining $G_p^{II,1}$ and $G_p^{II,0}$, we get our final result:
\bg\label{resone}
 G_p^{II}(\eta=0) &= & ~G_p^{II,0} ~ + ~ G_p^{II,1}\nonumber\\ 
& = & ~ \frac{H^2}{(2\pi)^3 2 p^3} \frac{H^2}{(2\pi)^2} \left[ 2 {\rm log} (\mu_{IR} / p)
+2{\rm log} (H/ {\widetilde\mu}_{UV})  \right] \, .
 \label{Gp2_d}
\nd

Similarly one may also compute $G_p^{I, 1}$. For this, 
if we are considering the UV divergence, we can use the 
approximation $\eta_2 \rightarrow \eta_1$. 
Since there are no IR divergences, this gives us: 
\bg\label{ade} 
G_p^{I,1}(\eta=0) =\frac{H^2}{(2\pi)^3 2 p^3} \frac{H^2}{(2\pi)^2} \left[   \frac{2}{3} {\rm log} (p/H) \right] 
\nd
which, when combined with $G_p^{I, 0}$ in \eqref{glynn}, gives us the following result:
\bg\label{loop2}
G_p^{I}(\eta=0) =\frac{H^2}{(2\pi)^3 2 p^3} \frac{H^2}{(2\pi)^2} \left[  2{\rm log} (p/\mu_{IR})  
+\frac{2}{3} {\rm log} ({\widetilde\mu}_{UV}/H) \right] 
\nd
One may now compare \eqref{resone} and \eqref{loop2} with \eqref{nhart} and \eqref{atexs} respectively that we got 
using brute cut-off. They match precisely for the logarithmic terms. One might however wonder about the 
additional quadratic piece in \eqref{nhart}:
\bg\label{abelfer}
-{H^2\over 2p^3 (2\pi)^5} \cdot {\widetilde \mu}^2_{UV}
\nd
This term cannot be seen from the dimensional regularization because this
method is optimised to capture the logarithmic divergences. Any divergences 
higher than logarithmic require a more involved regularization scheme.

\section{Pauli-Villars regularization}

Now that we have seen how the results from brute cut-off and dimension 
regularization match up precisely, it is 
time to analyse the UV/IR divergences using the Pauli-Villars regularization 
scheme. Our starting point is to 
take the interactions \eqref{2vert} alongwith a potential $V(\sigma)$ 
that could in principle appear from moduli stabilisation in M-theory. 

\subsection{Pauli-Villars regulators}

Pauli-Villars regularization is a scheme to cancel the divergences in loops 
by introducing one or a several massive fields, sacrificing  general 
covariance or gauge symmetry of the 
underlying theory. In this regularization process, 
the propagator of a field becomes proportional to: 
\bg\label{whstevn}
\frac{1}{k^2} - \frac{1}{k^2+M_0^2}= \frac{1}{k^2+k^4/M_0^2} \, ,
\nd
where $M$ is a cutoff mass. In the high energy $E >> M_0$ limit the 
propagator vanishes, cancelling the UV divergence 
by the heavy field. Similarly in the low energy $E << M_0$ limit the 
propagator regains its massless limit. 

The above is the standard story behind Pauli-Villars regularization. 
To extend this to more complicated scenarios 
we need to add more than one type of Pauli-Villars fields. This would, 
for example, change \eqref{whstevn} to the
following propagator:
\bg\label{bristevn} 
\frac{1}{k^2} +\sum_n\frac{Z_n^{-1}}{k^2+M_n^2}= \frac{1}{k^2 \prod_n (k^2+M_n^2)/ \prod_n M_n^2} \, ,
\nd
where $Z^{-1}_n$ are the typical coefficents for the free parts of the 
Pauli-Villars Lagrangian and $M_n$ are the masses of the regulator
fiels. These coefficients  satisfy:
\bg\label{bree}
\sum_{n} Z_n^{-1}=-1, ~~~~~~~~ \sum_n Z_n^{-1} M_n^2=0 \, .
\nd

To apply the Pauli-Villars regularization scheme to our case, we need 
two sets of Pauli-Villars fields: one set for 
the scalar field $\sigma$ and the other set for the graviton field $h_{ij}$. 
We will call these fields as $\chi_n$ and $\gamma^n_{ij}$ respectively. 
The typical lagrangian for our scalar field $\sigma$ in the presence of 
gravitational interaction is given by:
\bg\label{scallag} 
{\cal L} = \sqrt{-g} \left[{R\over 2} -\frac{a^2}{2}\partial_\mu \sigma  \partial^\mu \sigma -V_1(h_{ij}, \sigma)
-V_2(\sigma)\right] \, ,
\nd
where $V_2(\sigma)$ is a potential term for $\sigma$ whose form follows from
the ultraviolet theory being considered, and 
$V_1(h_{ij}, \sigma)$ is the minimal coupling of the metric 
$h_{ij}$ with $\sigma$ that generates the two interactions
\eqref{2vert}. 

Once we switch on the two sets of Pauli-Villars fields the 
Lagrangian with the scalar field $\sigma$ takes the following form:
\bg\label{lagPoV}
{\cal L}_s &= &\sqrt{-g} \Bigg[-\frac{a^2}{2}\partial_\mu \sigma  \partial^\mu \sigma 
-\frac{1}{2} \sum_n Z_n^{-1} \left(g^{\mu \nu} \partial_\mu \chi_n  \partial_\nu \chi_n  
+ M_n^2 \chi_n^2 \right)\nonumber\\
 &&  -V_1\left(h_{ij} + \sum_n \gamma_{ij}^n, \sigma + \sum_n \chi_n\right) - V_2\left(\sigma 
+ \sum_n \gamma_n\right)\Bigg] \, ,
\nd
where we have shifted the fields ($h_{ij}, \sigma$) by their corresponding 
regulator fields in the potential to generate the required couplings 
between them. The gravitational part of the Lagrangian can also be 
adjusted to take into account the action of the regulator fields:
\bg 
{\cal L}_g = \sqrt{-g} \left[\frac{1}{2} R_0 -\frac{1}{8} g^{\mu \nu} \partial_\mu h_{ij}  \partial_\nu h_{ij}  
-\frac{1}{8} \sum_n \widetilde{Z}_n^{-1} \left(g^{\mu \nu} \partial_\mu \gamma_{ij}^n  \partial_\nu \gamma_{ij}^n  
+ {\widetilde M}_n^2 {\gamma_{ij}^n}^2 \right)\right] \, ,
\nd
where $R_0$ is the curvature associated with the background 
de-Sitter spacetime and (${\widetilde Z}_n, {\widetilde M}_n$) satisfy 
relations similar to \eqref{bree}.
The above action for the graviton $h_{ij}$ and the regulator fields $\gamma^n_{ij}$ 
make sense because the free graviton field and free scalar field are the same up to 
some constant. Thus, adding 
the regulator fields should be identical for both cases. 

Having obtained the full action for our case, let us investigate the interaction terms. 
The linear shift in the potential $V_1(h_{ij}, \sigma)$ immediately gives us the 
following relevant interactions: 
\bg\label{lonmik}
&& {\cal L}_3 = \frac{a}{2} \left( h_{ij} + \sum_n \gamma_{ij}^n \right) \partial_i \sigma \partial_j \sigma\nonumber\\
&& {\cal L}_4 = -\frac{a}{4}  \left( h + \sum_n \gamma^n \right)_{il} 
\left( h + \sum_m \gamma^m \right)_{lj} \partial_i \sigma \partial_j \sigma \, .
\nd
The above are the two kinds of interactions that we will consider for our case. 
However there are additional interactions of the form:
\bg\label{addinter}
&& \frac{a}{2}\sum_{n,m,l} \left( h_{ij} + \gamma_{ij}^n \right)\left(\partial_i \sigma \partial_j \chi_m +  
\partial_i \chi_l \partial_j \chi_m\right) - V_2\left(\sigma + \sum_k \chi_k\right)\nonumber\\
&& ~~~~~~~~~ -{a\over 4} \sum_{n,m,k,p} \left( h + \gamma^n \right)_{il} \left( h + \gamma^m \right)_{lj} 
\left(\partial_i \sigma \partial_j \chi_k + \partial_i \chi_p \partial_j \chi_k\right) \, .
\nd

Before we analyse the above interactions, it is easy to see from \eqref{lagPoV}
that the interaction $V_2\left(\sigma + \sum_k \chi_k\right)$ leads to diagrams that are 
unambiguously regularized in \cite{Weinberg2}. 
So all we need to consider are the other interactions. Most of the interactions
in \eqref{addinter} are not one-loop. The only one-loop diagrams are given in Figure 4. 
Since both diagrams have the same $\chi_m$ fields propagating in the loop, the 
regularization for $h_{ij}$ proceeds in exactly the 
same way as in the case of the first interaction in \eqref{lonmik}. This is evident when 
we incorporate the diagrams in Figure 4 in our computations.

\begin{figure}[htb]\label{chichi}
\begin{center}
\includegraphics[height= 5cm]{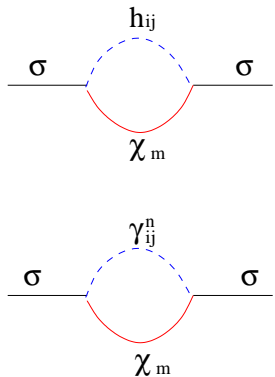}
\caption{The other one-loop interactions in the theory.}
\end{center}
\end{figure}

\subsection{Quantization of the Pauli-Villars fields}

As we see from \eqref{lonmik}, the only relevant Pauli-Villars fields that we need to 
consider in our case are the fields $\gamma_{ij}^n$, since the $\chi_n$ fields appear 
in diagrams that are not relevant in our case. Therefore, let us start 
by defining $\gamma_{ij}^0 \equiv h_{ij}$ with ${\widetilde Z}_0 \equiv 1$, 
${\widetilde M}_0 = 0$,  and $\gamma_{ij}^n$ for $n \ge 1$ being
the set of Pauli-Villars fields. We can then write down the following mode expansion:
\bg\label{asilver} 
\gamma^n_{ij} ({\bf z},\eta)
= \int d^3q \sum_{s=\pm} \left[a({\bf q}) e_{ij}(q, s) \gamma^n_{s,q}(\eta) e^{i{\bf q} 
\cdot {\bf z}} + a^\dagger({\bf q}) {e_{ij}}^* (q, s) \gamma^{n*}_{s,q}(\eta) e^{-i{\bf q} \cdot {\bf z}}\right]
\,
\nd
where $e_{ij}$ satisfy the identity \eqref{polT} and $\gamma^n_{s,q} \equiv \gamma^n_q$ 
denote the modes of $\gamma^n_{ij}$ for a given momentum $q$. The creation and 
annihilation operators then satify:
\bg\label{creannil} 
 [a_n({\bf k}_1),a_m^{\dagger}({\bf k}_2)] = \delta^3({\bf k}_1-{\bf k}_2)\delta_{mn} \, .
\nd
One may similarly quantise the other set of Pauli-Villars fields $\chi_n$, but will not do so 
here. The mode expansion of the fields $\chi_n$ is used to compute the two diagrams of Figure 4.

\subsection{The regularization process}

We are now ready to perform the actual regularization process using the 
two interactions \eqref{lonmik}. Our method will be very similar to the 
one recently developed in \cite{Weinberg2}, which the reader may consult for 
additional information. Note that the analysis of\cite{Weinberg2} 
only dealt with the scalar field $\sigma$ and its Pauli-Villars partners $\chi_n$. 
We therefore extend the technique of \cite{Weinberg2} to apply to graviton fluctuation.
  
As before, our aim is to regularize the two-point functions using a Pauli-Villars field. 
To do this, note that the internal momentum integral may be separated into two parts, 
exactly as in \cite{Weinberg2}. As one part we choose an interval where the internal 
momenta are much smaller than the masses of the regulator fields, in the other part 
we take the interval where the internal momenta are of the same order as the masses 
or larger. We will call the separation scale $Q$ (following \cite{Weinberg2}). It plays
no physical role. Therefore, the final two-point function should not depend on the scale Q. 

The two-point function for the second interaction in \eqref{lonmik} is then given
(using co-moving coordinates) by:
\bg\label{secdia}
G_p^{II} &=& \left(2 \pi \right)^3 \int_{-\infty}^\eta d\eta_1 a^2(\eta_1) {\bf Im} \Big[u_p^2(\eta){u_p^* }^2(\eta_1)
\Big] p_i p_j
\nonumber\\
&&
\times \sum_s \sum_{KMN} {\widetilde Z}^{-1}_K  \int d^3 q  e_{il}(\hat q,s)  e_{lj}^*(\hat q,s)
\gamma^K_{Mq}(\eta_1 ){\gamma^{K*}_{Nq}}(\eta_1 )  \, .
\nd
The above integral may now be divided into two intervals as discussed above
by introducing the momentum scale $Q$. If we also choose the 
masses ${\widetilde M}_n = M_n$ ($n \ge 1, {\widetilde M}_0 = 0$)
for simplicity, then we can express \eqref{secdia} for $q > Q$ in the following way:
\bg\label{secdiaQ1}
G_p^{II,>Q} &=& \frac{4}{3}\pi \int_{-\infty}^\eta d\eta_1 a^2(\eta_1) {\bf Im}\Big[u_p^2(\eta) {u_p^* }^2(\eta_1)
\Big]p^2 \nonumber\\
&& \left[\sum_{n} {\widetilde Z}_m^{-1} M_n^2 \ln M_n + \left(\dot H (\eta_1) +2 H^2(\eta_1)\right) 
\left(\ln 2 -\sum_n {\widetilde Z}_n^{-1} \ln M_n\right) \right.
\nonumber\\
&&  \left. -\frac{Q^2}{a(\eta_1)^2}- \left(\dot H (\eta_1) +2 H^2(\eta_1) \right) \ln \left(\frac{Q}{a(\eta_1)} \right)
\right]\nonumber\\
&=& \frac{4}{3}\pi \int_{-\infty}^\eta d\eta_1 a^2(\eta_1) {\bf Im} \Big[u_p^2(\eta) {u_p^* }^2(\eta_1)\Big] 
p^2 \nonumber\\
&& \left[\mu_A^2  + \left(\dot H (\eta_1) +2 H^2(\eta_1) \right)  {\rm log}  \left(  \frac{\mu_B a(\eta_1)}{Q} \right)  
-\frac{Q^2}{a(\eta_1)^2} \, ,
\right]
\nd
where in the second equality we have chosen the renormalisation scales $\mu_A$ and 
$\mu_B$. These two scales will be related to each other, but we will make the 
identification after we add the contributions from amplitudes with $q < Q$. 
For $q < Q$ our result is: 
\bg\label{kuttarb} 
G_p^{II,<Q}=  (2\pi)^3 \int_{-\infty}^{\eta} d\eta_1 a^2(\eta_1) {\bf Im} \Big[u_p^2(\eta) {u_p^* }^2(\eta_1)\Big] 
2p^2 \int_{\Lambda_{IR}}^Q d^3 q {\rm sin}^2 \theta 
\vert h_q(\eta_1) \vert^2 \, .
\nd
Once we restrict our result to the de-Sitter spacetime, 
the final amplitude will be given by the sum of the above two, i.e the sum of 
\eqref{secdiaQ1} and \eqref{kuttarb}. This gives
\bg\label{leahlu} 
G_p^{II} &\equiv & G_p^{II,>Q}+G_p^{II,<Q} \nonumber\\
&=& \frac{8}{3}\pi (2\pi)^3 \int_{-\infty}^\eta  d\eta_1 a^2(\eta_1) {\bf Im}\Big[u_p^2(\eta) {u_p^* }^2(\eta_1)\Big]p^2 
\nonumber\\
&& \left[\mu_A^2  + 2 H^2(\eta_1)  {\rm log}  \left(\frac{\mu_B a(\eta_1)}{\Lambda_{IR}} \right) 
\right]\nonumber\\
&=& \frac{H^2}{(2\pi)^3 2 p^3} \frac{H^2}{(2\pi)^2} \left[ 2 \log (\mu_{IR} / p) 
+ 2\log (H/{\widetilde \mu}_{UV}) - \frac{{\widetilde \mu}_{UV}^2}{H^2} \right] \, ,
\nd
where to go from the second step to the final one, we have identified 
$\mu_A = \mu_B = {\widetilde \mu}_{UV}$ and 
$\Lambda_{IR} \equiv \mu_{IR} a$. This matches precisely with  
the other two regularization schemes defined with 
physical UV cut-off and co-moving IR cut-off. 

Finally, for the first interaction of \eqref{lonmik} the amplitude is give by the following expression 
(again using co-moving coordinates):
\bg\label{jhutamaro}
G_p^{I} &=& -4\left(2 \pi \right)^6 {\bf Re}~ \int_{-\infty}^\eta d\eta_1 a^2(\eta_1) \int_{-\infty}^{\eta} d\eta_2
 a^2(\eta_2) \nonumber\\
 && \times \sum_s \sum_{KLMNM'N'} {\widetilde Z}^{-1}_{K} Z_L^{-1}  \int d^3 q ~ p_i p'_j p_k p'_l~ e_{ij}(\hat q,s)  
e_{kl}^*(\hat q,s)\nonumber\\
 &&\times \Bigg[\theta(\eta_1 - \eta_2) u_p^2(\eta) u_p^*(\eta_1) u_p^*(\eta_2) \gamma^K_{Mq}(\eta_1 ){\gamma^{K*}_{M'q}}
(\eta_2 ) u^L_{Np'}(\eta_1)
{u^{L*}_{N'p'}}(\eta_2) \nonumber\\
&& -\frac{1}{2} \vert u_p(\eta)\vert^2 u_p^*(\eta_1) u_p(\eta_2)\gamma^{K*}_{Mq}(\eta_1 ) \gamma^{K}_{M'q}(\eta_2) 
u^{L*}_{Np'}(\eta_1) u^{L}_{N'p'} (\eta_2)\Bigg] \, .
 \nd
The above integral can again be restricted to the two intervals exactly as before. For $q > Q$, \eqref{jhutamaro} takes the following form:
\bg\label{chappalmaro}
G_p^{I,>Q} &=& \frac{16}{15}\pi \int_{-\infty}^\eta d\eta_1 a^2(\eta_1) {\bf Im} \Big[u_p^2(\eta) {u_p^* }^2(\eta_1)
\Big]p^4 \nonumber\\
&& \left[\sum_{mn} {\widetilde Z}_m^{-1} Z_n^{-1} \frac{M_n^2 \ln M_n -M_m^2 \ln M_m }{M_n^2-M_m^2} 
+ \sum_n {\widetilde Z}_n^{-1} \ln M_n
+\sum_n Z_n^{-1} \ln M_n +\ln \left( \frac{Q}{a(\eta_1)} \right)\right]\nonumber\\
&=& \frac{16}{15}\pi \int_{-\infty}^\eta d\eta_1 a(\eta_1) {\bf Im} \Big[u_p^2(\eta) {u_p^* }^2(\eta_1)\Big] 
p^4  {\rm log} \frac{Q}{a(\eta_1) {\widetilde \mu}_{UV}} \, ,
\nd
where ${\widetilde \mu}_{UV}$ is the required physical UV renormalisation scale. On the other hand,
for $q < Q$ we have the following expression which is a slight variation of \eqref{Gp1_c}:
\bg\label{laddu}
G_p^{I,<Q} &=& -4 (2\pi)^6 {\bf Re} \int_{-\infty}^{\eta}d\eta_1 a^2(\eta_1)  
\int_{-\infty}^{\eta}d\eta_2 a^2(\eta_2) 
\int_{\Lambda_{IR}}^Q d^3 q\  p^4 {\rm sin}^4 \theta \nonumber\\
 &&\times \Big[\theta (\eta_1-\eta_2) u_p^2(\eta) u_p^*(\eta_1) u_p^*(\eta_2)  u_{p'}(\eta_1) u_{p'}^*(\eta_2) 
h_q(\eta_1) h_q^*(\eta_2)\nonumber\\
 &&-\frac{1}{2}  |u_p(\eta)|^2  u_p(\eta_1) u_p^*(\eta_2)  u_{p'}(\eta_1) u_{p'}^*(\eta_2) h_q(\eta_1) 
h_q^*(\eta_2)\Big] \, .
\nd   
In de-Sitter space we can add up \eqref{laddu} and \eqref{chappalmaro} to give us the 
following final expression for the one-loop interaction:
\bg\label{atxas} 
G_p^{I} &\equiv & G_p^{I,>Q} ~ + ~ G_p^{I,<Q}\nonumber\\
&=&\frac{H^2}{(2\pi)^3 2 p^3} \frac{H^2}{(2\pi)^2} \left[2\log (p/\mu_{IR})  
+\frac{2}{3} \log ({\widetilde\mu}_{UV}/H) \right] \, ,
\nd
which again matches precisely with the results that we got from the other two regularization schemes.  

\section{Conclusions}

In this paper we have computed one loop graviton corrections to the Green's function
of a scalar matter field in de Sitter space using three different regularization schemes: 
brute-force cutoff, dimensional regularization and the Pauli-Villars prescription. 
We have shown that careful evaluation in the three cases leads to the identical result. 

There are both infrared and ultraviolet divergences which appear in the one loop
computation. By embedding the de Sitter phase into an ultra-violet complete theory
we were able to justify the use of an ultraviolet cutoff at a fixed physical scale.
This leads to the conclusion that the Hilbert space of modes of the effective low energy
field theory is growing as space is expanding, and this in turn leads to the presence
of a growing contribution of infrared modes to correlation functions. 

The embedding of an inflationary model in the context of string theory allows us
to study trans-Planckian ``problem" \cite{TPproblem} for cosmological fluctuations 
in inflationary cosmology. In toy models of inflation, it is unclear what state to choose
for the fluctuation modes when they arise from the ``trans-Planckian sea". The
answer, however, will be well defined once the inflationary model is embedded in
an ultraviolet complete theory. We are currently investigating this issue \cite{toappear3}. We
are also planning to study further consequences of the growth of the phase space
of infrared modes which contribute to correlation functions.

\section*{Acknowledgements}

This work is supported in part by NSERC Discovery grants to R.B. and K.D. and by funds from the
Canada Research Chair program (R.B.). W.X. is supported by a Schulich fellowship, and R.B. is 
recipient of a Killam Research Fellowship. We would like to thank Yifu Cai, Alex Maloney, 
Guy Moore, Sachin Vaidya, and Yi Wang for helpful discussions, and S. Giddings, L. Senatore, M. Sloth and in particular R. Woodard for comments on the draft. Our interest in this topic was 
stimulated by the Perimeter Institute workshop on ``IR Issues and Loops in de Sitter Space" held 
October 27 - 30, 2010. We thank C. Burgess, R. Holman, L. Leblond and S. Shandera
for organizing this interesting workshop.  


\end{document}